\input harvmac
\newcount\figno
\figno=0
\def\fig#1#2#3{
\par\begingroup\parindent=0pt\leftskip=1cm\rightskip=1cm\parindent=0pt
\global\advance\figno by 1
\midinsert
\epsfxsize=#3
\centerline{\epsfbox{#2}}
\vskip 12pt
{\bf Fig. \the\figno:} #1\par
\endinsert\endgroup\par
}
\def\figlabel#1{\xdef#1{\the\figno}}
\def\encadremath#1{\vbox{\hrule\hbox{\vrule\kern8pt\vbox{\kern8pt
\hbox{$\displaystyle #1$}\kern8pt}
\kern8pt\vrule}\hrule}}

\overfullrule=0pt

%
\def\underarrow#1{\vbox{\ialign{##\crcr$\hfil\displaystyle
 {#1}\hfil$\crcr\noalign{\kern1pt\nointerlineskip}$\longrightarrow$\crcr}}}
%
\def\tilde{\widetilde}
\def\bar{\overline}

\def\inbar{\vrule height1.5ex width.4pt depth0pt}
\def\IC{\relax\hbox{\kern.25em$\inbar\kern-.3em{\rm C}$}}
\def\IR{\relax\hbox{\kern.25em$\inbar\kern-.3em{\rm R}$}}
\def\IZ{\relax\ifmmode\hbox{Z\kern-.4em Z}\else{Z\kern-.4em Z}\fi}

\font\zfont = cmss10 

\def\bigone{\hbox{1\kern -.23em {\rm l}}}
\def\ZZ{\hbox{\zfont Z\kern-.4emZ}}


\def\drawbox#1#2{\hrule height#2pt
        \hbox{\vrule width#2pt height#1pt \kern#1pt
              \vrule width#2pt}
              \hrule height#2pt}

\def\Asym#1#2{\vcenter{\vbox{\drawbox{#1}{#2}
              \kern-#2pt       
              \drawbox{#1}{#2}}}}

\batchmode
  \font\bbbfont=msbm10
\errorstopmode
\newif\ifamsf\amsftrue
\ifx\bbbfont\nullfont
  \amsffalse
\fi
\ifamsf
\def\IR{\hbox{\bbbfont R}}
\def\IC{\hbox{\bbbfont C}}

\def\IZ{\hbox{\bbbfont Z}}


\midinsert
\endinsert


\nref\polchinski{ J. Polchinski, ``TASI Lectures on D-branes'',
in
{\it Fields Strings and Duality} TASI-96, Eds. C. Efthimiou and B. Greene, World
Scientific (1997), hep-th/9611050.}

\nref\michel{M. Douglas, ``Superstring Dualities, Dirichlet Branes and the
Small Scale Structure of Space'', hep-th/9610041.}

\nref\gepner{D. Gepner and E. Witten, ``String Theory on Group Manifolds'',
Nucl. Phys. B {\bf 278} (1986) 493-549.}

\nref\severa{ C. Klim$\check{\rm c}$ik and P. $\check{\rm S}$evera,
``Poisson-Lie T-duality: Strings and D-branes'', Phys. Lett. B {\bf 376}
(1996) 82; ``Open Strings and D-branes in WZNW Models'', Nucl. Phys. B {\bf
488} (1997) 653-676.}

\nref\kato{ M. Kato and T. Okada, ``D-branes on Group Manifolds'', Nucl. Phys. 
B {\bf 499}  (1997) 583, hep-th/9612148.}

\nref\stanciu{S. Stanciu. ``D-branes in Kazama-Susuki Models'',
Nucl. Phys. B {\bf 526} (1998) 295, hep-th/9708166.}

\nref\reck{ A. Recknagel and  V. Schomerus, ``D-branes in Gepner Models'',
Nucl. Phys. B {\bf 531} (1998) 185, hep-th/9712186.}

\nref\yu{ A. Yu. Alekseev and V. Schomerus, ``D-branes in the WZW Model'',
hep-th/9812193.}

\nref\gawedzki{ K. Gawedzki, ``Conformal Field Theory: A Case Study'',
 hep-th/9904145.}
 
\nref\bayen{F. Bayen, M. Flato, C. Fronsdal, A. Lichnerowicz and D. Sternheimer, 
``Deformation Theory and Quantization I;II'', Ann. Phys. {\bf 111} (1978) 61;111; M. De 
Wilde  and P.B.A. Lecomte, Lett. Math. Phys. {\bf 7} (1983) 487; H. Omory, Y. Maeda
and A. Yoshioki, ``Weyl Manifolds and Deformation Quantization'', Adv. Math. {\bf 85} (1991)
224; B. Fedosov, ``A Simple Geometrical Construction of Deformation Quantization'', J. 
Diff. Geom. {\bf 40} (1994) 213.}
 
\nref\k{M. Kontsevich, ``Deformation Quantization of Poisson
Manifolds I'', q-alg/9709040.}

\nref\bara{S. Barannikov and M. Kontsevich, ``Frobenius Manifolds and
Formality of Lie Algebras of Polyvector Fields'', alg-geom/9710032.}

\nref\gutt{ S. Gutt, ``An Explicit $*$-product on the Cotangent Bundle of
a Lie Group'', Lett. Math. Phys. {\bf 7} (1983) 249-258.}

\nref\asin{S. Asin Lares, ``On Tangential Properties of the Gutt
$*$-Product'', J. Geom. Phys. {\bf 24} (1998) 164.}

\nref\kathotia{ V. Kathotia, ``Kontsevich's Universal Formula For
Deformation Quantization and the Campbell-Baker-Hausdorff Formula, I'',
math.QA/9811174.}

\nref\dito{G. Dito, ``Kontsevich Star-product on the Dual of a Lie
Algebra'', math.QA/9905080.}

\nref\jerzy{B. Mielnik and J.F. Pleba\'nski, ``Combinatorial approach to
Baker-Campbell-Hausdorff Exponents'' Ann. Inst. Henri Poincar\'e {\bf 12}
(1970) 215-254.}

\nref\sho{B. Shoikhet, ``On the Kontsevich and the CBH Deformation
Quantization of a Linear Poisson Structure'', math.QA/9903036.}

\nref\kaz{P. Etingof and D. Kazhdan, ``Quantization of Lie bialgebras, I;II;III'',
Sel. math., new ser. {\bf 2} (1996) 1; {\bf 4} (1998) 213; {\bf 4} (1998) 233.}

\nref\tamarkin{D. Tamarkin, ``Another Proof of M. Kontsevich Formality Theorem'',
math.QA/9803025.}

\nref\maxim{M. Kontsevich, ``Operands and Motives in Deformation
Quantization'', math.QA/9904055.}

\nref\cattaneo{A.S. Cattaneo and G. Felder, ``A Path Integral Approach
to the Kontsevich Quantization Formula'', math.QA/9902090.}

\nref\schwarz{ A. Schwarz, ``Quantum Observables, Lie Algebra Homology
and TQFT'', hep-th/9904168.}

\nref\arefeva{ I.Ya. Aref'eva and I.V. Volovich, ``Noncommutative Gauge Fields
on Poisson Manifolds'', hep-th/9907114.} 

\nref\schomerus{ V. Schomerus, ``D-branes and Deformation Quantization'',
JHEP {\bf 06} ((1999) 030, hep-th/9903205.}

\nref\cds{A. Connes, M.R. Douglas and A. Schwarz, ``Noncommutative Geometry and
Matrix Theory: Compactification on Tori'', JHEP {\bf 02} (1998) 003, hep-th/9711162.}

\nref\douglas{M. Douglas, `` Two Lectures on D-Geometry and 
Noncommutative Geometry'', hep-th/9901146.}

\nref\arda{F. Ardalan, H. Arfaei and M.M. Sheikh-Jabbari, ``Dirac Quantization of Open
Strings and Noncommutativity in Branes'', hep-th/9906161.}

\nref\ho{Chong-Sun Chu and Pei-Ming Ho, ``Constrained Quantization of Open Strings in
Background $B$-field and Noncommutative D-brane'', hep-th/9906192.}

\nref\hugo{H. Garc\'{\i}a-Compe\'an, ``On the Deformation Quantization
Description of Matrix Compactification'', Nucl. Phys. B {\bf 541} (1999)
651-670.}

\nref\ed{E. Witten, ``Nonabelian Bosonization in Two Dimensions'', Commun. Math. 
Phys. {\bf 92} (1984) 455-472.}

\nref\branes{E. Witten, ``Bound States of Strings and $p$-Branes'', Nucl. Phys. B
{\bf 460} (1996) 335, hep-th/9510135.}

\nref\cs{E. Witten, ``Chern-Simons Gauge Theory as a String Theory'',
 hep-th/9207094.}

\nref\fioresi{R. Fioresi and M.A. Lled\'o. ``On the Deformation
Quantization of Coadjoint Orbits of Semisimple Groups'',
math.Q.A./9906104.}

\nref\vafa{M. Bershadsky, C. Vafa and V. Sadov, ``D-branes and
Topological QFT's'', Nucl. Phys. B {\bf 463} (1996) 420.}

\nref\kz{E. Witten, ``The $N$-matrix Model and Gauged WZW Models, Nucl. Phys. B 
{\bf 371} (1992) 191.}

\nref\roza{L. Rozansky and E. Witten, ``Hyper-K\"ahler Geometry and Invariants for
Three-manifolds'', Sel. math., new ser. {\bf 3} (1997) 401.}

\nref\recknagel{ A. Recknagel and  V. Schomerus, ``Boundaries Deformation Theory
and Moduli Spaces of D-branes'', Nucl. Phys. B {\bf 545} (1999) 233, 
hep-th/9811237.}

\nref\bordemann{ M. Bordemann, N. Neumaier, S. Waldmann, Commun. Math. Phys. {\bf 198}
(1998) 363.}




\rightline{hep-th/9907183, IASSNS-HEP-99/70}
\Title{CINVESTAV-FIS-99/33}
{\vbox{\centerline{D-branes on Group Manifolds}
\medskip
\centerline{and Deformation Quantization  }}}
\smallskip
\centerline{Hugo
Garc\'{\i}a-Compe\'an$^{a,b}$\foot{E-mail:
compean@fis.cinvestav.mx} and Jerzy F. 
Pleba\'nski$^b$\foot{E-mail: pleban@fis.cinvestav.mx}}
\smallskip
\centerline{\it $^a$School of Natural Sciences}
\centerline{\it Institute for Advanced Study}
\centerline{\it Olden Lane, Princeton, NJ 08540, USA}
\smallskip
\centerline{\it $^b$Departamento de F\'{\i}sica}
\centerline{\it  Centro de Investigaci\'on y de Estudios Avanzados del IPN}
\centerline{\it Apdo. Postal 14-740, 07000, M\'exico D.F., M\'exico}

\bigskip
\medskip
\vskip 1truecm
\noindent

Recently M. Kontsevich found a combinatorial formula defining a star-product
of deformation quantization for any Poisson manifold. Kontsevich's formula
has been reinterpreted physically as quantum correlation functions of a
topological sigma model for open strings as well as in the context of
D-branes in flat backgrounds with a Neveu-Schwarz $B$-field.  Here the
corresponding Kontsevich's formula for the dual of a Lie algebra is
derived in terms of the formalism of D-branes on group manifolds.  In
particular we show that that formula is encoded at the two-point
correlation functions of the Wess-Zumino-Witten effective theory with Dirichlet
boundary conditions. The $B$-field entering in the formalism plays an
important role in this derivation.


\noindent

\Date{July, 1999}



\newsec{Introduction}

D-branes are extremely interesting objects which were 
originally applied to realize string dualities and to probe substringly distances
(for a review, see Ref.  \polchinski). One of the most exciting aspects of
D-brane physics is the realization and explanation of several mathematical
constructions in the language of physical processes of branes. For instance, the ADHM
construction of Yang-Mills instantons founds a nice realization in terms of
D-brane configurations (for a review of this subject, see Ref. \michel). 

Integrability present in the formulation of the string theory on group
manifolds, originally studied in \gepner, provided an excellent guide in
the search of a realistic string theory. The study of D-branes on group
manifolds was first considered by Klim$\check{\rm c}$ik and $\check{\rm
S}$evera in \severa\ and further developed in \refs{\kato,\stanciu,\reck}
from the point of view of boundary states theory. Some results of Ref.
\severa\ will be of particular usefulness for the purposes of this paper
and they will be reviewed at the section 3. Of special interest is the
result of \severa\ that a WZW Lagrangian can be well defined for open strings
propagating in a group manifold.  This effective Lagrangian can be
obtained from the pairing between the relative homology chain and the
relative cohomology chain, {\it i.e.} $H_*({\bf G}/{\bf H}_m,\IZ)  \otimes
H^*({\bf G}/{\bf H}_m,\IZ)\to \IZ$ where the integer-valued relative homology and cohomology
ensures a well defined Feynman path integral. Here ${\bf G}$ is the group manifold and
${\bf H}_m$ is a group submanifold of ${\bf G}$. Within this
context the D-branes of various dimensionalities are identified with the
submanifolds ${\bf H}$ and the open world-sheet $W$ generated b`y the
propagation of the open string in ${\bf G}$ can be identified with the
relative homology cycles of the corresponding dimension.  More recently,
it has been shown that D-branes identified with the submanifolds ${\bf
H}_m$ correspond to the conjugacy classes of the relevant Lie group ${\bf
G}$.  Thus open strings end points are permitted to be fixed in such a
classes and stretched between pairs of them \yu\ (for a very nice recent
review on the subject, see Ref. \gawedzki). 

On the other hand very recently a renewed deal of excitation has been taken
place in deformation quantization theory \bayen, since
the Kontsevich's seminal paper \k. In this paper Kontsevich proved by
construction the existence of a star-product for any finite dimensional
Poisson manifold. His construction is based on his more general statement
known as the ``formality conjecture''. The existence of such a
star-product determines the existence of a deformation quantization for
any Poisson manifold. Kontsevich's proof was strongly motivated by some
perturbative issues of string theory and topological gravity in two-dimensions,
such as, matrix models, the triangulation of the moduli space of Riemann surfaces
and  mirror symmetry \bara. 

Kontsevich's star product can also be defined for the dual of a Lie
algebra ${\cal G}^*$. The structure of ${\cal G}^*$ as  vector space it is 
known to be isomorphic to $\IR^n$,
being $n$ the number of generators of the Lie algebra ${\cal G}$. ${\cal
G}^*$ with the Kirillov-Poisson structure is a Poisson manifold and the
Kontsevich's star-product can be defined \k. It is also well known that other star-products
can be defined on ${\cal G}^*$ for any Lie algebra ${\cal G}$. The first of them was 
obtained by Gutt \gutt\ by studying the Hochschild cohomology of Lie algebras. There it was 
used not the Kirillov-Poisson structure on ${\cal G}^*$ but only the symplectic structure on
the cotangent bundle to the relevant Lie group (for a recent progress on this subject, 
see \asin). Gutt's star-product is naturally connected to the termed Campbell-Baker-Hausdorff
(CBH) quantization through suitable deformations of the universal enveloping algebras as we
shall review in Section 2 following Refs. \refs{\kathotia,\dito}.

The relation of Kontsevich's
quantization to the CBH quantization, was amply discussed in
\kathotia\ (see also \dito). It was shown in these references that both the Kontsevich's and 
universal enveloping algebra quantization (also called ${\cal U}$-quantization), are obtained 
from the exponentiation of the Kirillov-Poisson structure on ${\cal G}^*$, through the 
CBH-formula (for a review about the CBH-formula for the discrete and continuos cases, 
see \jerzy). Moreover both quantizations are exactly the same when Kontsevich's formula
is restricted to consider only symmetric-admissible graphs (with the loop-graphs being
neglected \kathotia). In Ref. \sho, such equivalence is further
investigated from the point of view of the formality conjecture. Finally the study of
Lie bi-algebras and their quantization was initialized  by Drinfeld. He conjectured 
that every Poisson-Lie group has a canonical quantization. The proof of this conjecture was 
given recently by Etingof and Kazhdan \kaz. Recently a new proof of the formality conjecture was
given by Tamarkin \tamarkin\ involving operads and motives. The implications of this new proof in 
the deformation quantization theory was more recently given by Kontsevich \maxim.

Afterwards a path integral interpretation of the Kontsevich's star-product
was worked out very recently  by Cattaneo and Felder \cattaneo. In this paper
Kontsevich's formula is extracted from the three-point correlation
functions on the disc from the perturbation theory of a bosonic
topological open string theory on the target space  provided with the relevant
Poisson structure.  Formality conjecture issues were also described there
in field theoretic terms. Further developments in this context were
worked out at \schwarz. The study of noncommutative gauge theories on Poisson manifolds
was recently discussed in \arefeva.

Among various applications of D-branes of particular interest, for the
purposes of this paper, is that of the D-brane realization of the
Kontsevich's star-product \k. Kontsevich's star product can be also
derived from the correlation functions of open string theory in for flat
backgrounds, but with non-trivial constant $B$-field \schomerus.  In this paper we
pursue a similar derivation of the Kontsevich's quantization formula for the
dual of a Lie algebra in field theoretic terms. In particular we show that
it can be obtained from the Wess-Zumino-Witten theory for open strings with Dirichlet
boundary conditions.  To be more precise we will show that that Kontsevich
star product is encoded in the effective field theory on the D-brane in group manifolds
described in \severa.

One of the main lessons of the stringly \cattaneo\ and D-brane \schomerus\ descriptions
of Kontsevich's formula in that of the deformation quantization for any Poisson
manifold requires necessarily of string theory. The deformation parameter of this
quantization is precisely the string scale $\alpha '$ (or the string coupling constant)
which in the limit $\alpha ' \to 0$ it reproduces the field theory limit but in this
limit the deformation quantization does not exist. The deformation arising precisely
when $\alpha ' \not= 0$ is an indication that deformation quantization is an stringly
phenomenon. Actually it was already suspected since the origin of the formality
conjecture where several mathematical ingredients of string theory were present.

Deformation quantization also arises in the description of the behavior
of D-branes in type IIA and IIB superstring theory on tori and in a background
$B$-field \cds. In particular, the M(atrix)-theory compactifications on tori have
been seen described through a Yang-Mills gauge theory (with sixteen
supercharges) on the noncommutative dual tori (for a recent review, see
\douglas\ and for some recent progress, see \refs{\arda,\ho}). Fedosov deformation quantization can be
also incorporated into
this context following the lines of \hugo.

We start in Section~2 by reviewing some basic features of the relevant star-products of
the dual Lie algebra ${\cal G}^*$, we will follow mainly the work of Kathotia
\kathotia. In Section~3 we describe some necessary facts about the theory of D-branes
on group manifolds \severa. In Section~4 we discuss the derivation of the Kontsevich's
formula for ${\cal G}^*$ from the theory of D-branes on group manifolds. Finally,
Section~5 contains our concluding remarks.

\vskip 2truecm


\newsec{Overview of the Various Star-Products in  the Dual of a Lie Algebra}

In this section we shall give an overview of the star-products on
the dual Lie algebra ${\cal G}^*$, in particular of the Kontsevich's star product
and its relation with the various star-products arising in the
literature. Our aim is not to provide an extensive review of such
constructions, but to briefly recall the relevant structure of these
quantizations and provide the notation, which we will need in the following
sections. For a more complete treatment see Refs. \refs{\kathotia,\dito,\sho}.
We will follow mainly the notation of Ref. \kathotia.

\vskip 1truecm

\subsec{Linear Poisson Structure. ${\cal U}$-quantization}

Let us first briefly review the linear Poisson structures. For
definiteness we are going to consider the deformation quantization of the dual Lie
algebra ${\cal G}^*$ to the Lie algebra ${\cal G}$ associated to a compact
and simple Lie
group ${\bf G}$. ${\cal G}^*$ is provided with the canonical Kirillov-Poisson structure given by
the structure constants of ${\cal G}$, $f^{ij}_{k}$ in the way $[X^i,X^j]=
f^{ij}_k X^k$, for any pair of elements $X^i$ and $X^j$ of a basis of ${\cal G}$. The
Kirillov-Poisson structure on ${\cal G}^*$ is given by $\alpha = {1\over 2}
f^{ij}_k X^k \partial_i \wedge \partial_j$. Here the elements $X^i$ can be seen also as
the local basis on ${\cal G}^*$. Now in order to induce a star-product on
the space of smooth functions $C^{\infty}({\cal G}^*)$ on ${\cal G}^*$,
one has to consider a suitable deformation of 
the universal
enveloping algebra ${\cal U}({\cal G}_{\hbar})$ and the deformation of the
symmetric tensor algebra
${\cal S}({\cal G})[[\hbar]]$ of ${\cal G}$ where $\hbar$ is the deformation parameter.  
${\cal S}({\cal G})[[\hbar]]$ can be
identified with the space of polynomial functions on ${\cal G}$ taking values in the
infinite formal series in $\hbar$. Thus if
$\sigma$ is the algebra isomorphism (given by the symmetrization of monomials
through the Poincar\'e-Birkhoff-Witt theorem)
between ${\cal U}({\cal G}_{\hbar})$  and ${\cal
S}({\cal G})[[\hbar]]$, it can be induced a star-product on $C^{\infty}({\cal G}^*)
[[\hbar]]$ in the form $P \star Q = \sigma^{-1} \big( \sigma(P) 
\circ \sigma(Q) \big)$, where $P,Q$ are two elements on $C^{\infty}({\cal
G}^*)[[\hbar]]$ and $\circ$ is an associative and noncommutative product in
the universal enveloping algebra ${\cal U}$\foot{ In the above procedure it was used the
fact that $C^{\infty}({\cal G}^*)[[\hbar]]$ and ${\cal S}({\cal G})
[[\hbar]]$ are canonically isomorphic.}. Linear Poisson structure of ${\cal G}^*$
leads naturally to two equivalent quantizations, the ${\cal
U}-$quantization and the CBH quantization. For
the purposes of this paper is enough to describe more in detail, the CBH-quantization.
The star product defined as above is equivalent to the Gutt's
star-product defined on the cotangent bundle to the Lie group ${\bf G}$ \gutt. Of course
they coincide with the star-product in the direction of the Kirillov-Poisson $\alpha$,
that is, $P \star Q = {1 \over 2} f^{ij}_k X^k {\buildrel{\leftarrow} \over \partial_i} P
\cdot 
{\buildrel{\rightarrow} \over {\partial}_jQ}.$ Thus the star-product on ${\cal G}^*$ can 
generated through the CBH-formula. This is the subject of the next subsection.

\vskip 1truecm

\subsec{Campbell-Baker-Hausdorff Quantization}

The Campbell-Baker-Hausdorff quantization is based in the CBH-formula
for the group multiplication. The product of two group elements of the form
${\rm exp}(X)$ and ${\rm exp}(Y)$  with $X,Y$ being elements of the Lie algebra ${\cal G}$ is given by

\eqn\uno{{\rm exp}(X) \cdot {\rm exp}(Y) = {\rm exp}(H(X,Y)),}
where $H(X,Y)$ is a series of the form
\eqn\dos{H(X,Y) = X+Y + {1\over 2} [X,Y] + {1 \over 12} ([X,[X,Y]] + [X,Y],Y]) +
\dots}
with $[X,Y]$ is the Lie bracket in ${\cal G}$.

Now we consider a basis of ${\cal G}$ given by $\{X^i \}$ with $i= 1,
\dots , d$. By using the Kirillov-Poisson structure, the generators of ${\cal G}$ are 
also seen as coordinates in ${\cal G}^*$. This permits to establish, for any small real
number $t$, the correspondence between the functions ${\rm exp}(tX)$ on ${\cal
G}^*$ and the group elements of the form
${\rm exp}(tX)$ with $X \in {\cal G}$

\eqn\tres{{\rm Functions} \ \ {\rm on} \ \  {\cal G}^* \Longleftrightarrow 
{\rm Group} \ \ {\rm  elements} \ \ 
{\rm exp}(tX) \ \ {\rm with} \  X \in {\cal G}.}
Pulling-back the group multiplication to the functions via this
correspondence leads to the CBH-quantization. This correspondence will be
of prime importance below in the construction of the star-product through the 
D-brane technology in section 4. Meanwhile we will describe this correspondence with
certain detail. Under this correspondence the CBH-formula induces a
bi-differential operator $\hat{\cal D}: C^{\infty}({\cal G}^*) \otimes C^{\infty}({\cal G}^*)
\to C^{\infty}({\cal G}^*)$. The procedure to
construct this operator is by constructing first its symbol $D$ and then
finding out its corresponding operator. Consider first the relation

\eqn\tres{ {\rm exp}\bigg(H(X,Y) -X-Y\bigg) = {\rm exp}\bigg({1 \over 2} [X,Y] + {1\over 12}
([X,[X,Y]] + [[X,Y],Y]) + \dots \bigg) }
coming from the group multiplication of the elements ${\rm exp}(X)$ and ${\rm exp}(Y)$.
Substituting $X=s_iX^i$ and $Y = t_j X^j$ in the above formula, its rhs provides a 
definition of the symbol $D$ of the operator $\hat{\cal D}$ given by

\eqn\cuatro{ D \equiv {\rm exp} \bigg( {1 \over 2} X^k f^{ij}_k s_it_j + {1 \over 12} (X^k
f^{im}_kf^{lj}_m s_i s_l t_j + X^k f^{ml}_k f^{ij}_m s_i t_j t_l) + \dots
\bigg)}
where $s_i$ and $t_i$ $i,j=1, \dots , d$ are real valued commuting variables.

This formula tell us that the symbol $D$ depends linearly on  the variables $X^k$ with
$X^k$ being  the local coordinates  on ${\cal G}^*$ which are commuting variables too. Thus
in the definition of the symbol given by Eq. \cuatro\ there is no noncommuting variables which
leads to some ambiguities.
The operator $\hat{\cal D}$ by itself
is determined as usual by the substitution of $s_i \leftrightarrow
\buildrel{\leftarrow}\over{\partial}_i$ and $t_j \leftrightarrow \buildrel{\rightarrow}
\over{\partial}_j$. It is given by

\eqn\cinco{\hat{\cal D}= {\rm exp}\bigg( {1 \over 2} X^k f^{ij}_k 
\buildrel{\leftarrow}\over {\partial_i} \cdot \buildrel{\rightarrow}\over {\partial_j}
 + {1 \over 12} \big(X^k
f^{im}_kf^{lj}_m \buildrel{\leftarrow}\over{\partial}_i \buildrel{\leftarrow}\over{\partial}_l
\buildrel{\rightarrow} \over{\partial}_j
+ X^k f^{ml}_k f^{ij}_m \buildrel{\leftarrow}\over{\partial}_i
\buildrel{\rightarrow} \over{\partial}_j \buildrel{\rightarrow} \over{\partial}_l\big)
+ \dots \bigg).}

Thus $\hat{\cal D}$ determines the {\it associative} star-product on
$C^{\infty}({\cal G}^*)$ given by

\eqn\seis{F \star G \equiv \hat{\cal D}(F,G)}
for the functions $F= {\rm exp}(sX)$ and $G= {\rm exp}(tY)$ on ${\cal G}^*$. This product 
determines a deformation of the point-wise multiplication
in the direction of the Poisson bracket $\alpha$.

The associativity of $\star$ is induced from the associativity of the group multiplication
via the pull-back  of such a property from the CBH-formula. Written explicitly Eq. \seis\ looks
like

\eqn\siete{ {\rm exp}(s_iX^i) \star {\rm exp}(t_jX^j) =
\hat{\cal D} \bigg({\bf X}, {\bf f}, (\buildrel{\leftarrow}\over {\partial_1}, \dots
, \buildrel{\leftarrow}\over {\partial_d}),(\buildrel{\rightarrow}\over {\partial_1},
\dots , \buildrel{\rightarrow}\over {\partial_d})\bigg)\bigg({\rm exp}(s_iX^i), {\rm
exp}(t_jX^j)\bigg)}
where we have explicitly displayed the  dependence  of the bi-differential operator
$\hat{\cal D}$ on the vector field
${\bf X}$, the structure constants ${\bf f}$ and the higher order derivatives.

In terms of the symbol $D$ this last expression is given by

\eqn\nueve{ {\rm exp}(s_iX^i) \star {\rm exp}(t_jX^j) =
D \bigg({\bf X}, {\bf f}, (s_1,s_2, \dots , s_d),(t_1,t_2, \dots t_d)\bigg)
{\rm exp}(s_iX^i){\rm exp}(t_jX^j)}
where $D$ is given by the Eq. \cuatro. For future convenience we write another form for 
the 
star-product \nueve\ 
$$
{\rm exp}(s_iX^i) \star {\rm exp}(t_jX^j) =
{\rm exp}\bigg(H(s_iX^i,t_jX^j) -s_iX^i-t_jX^j\bigg) {\rm exp}(s_iX^i) {\rm exp}(t_jX^j)$$

\eqn\diez{=  {\rm exp}\big(H(s_iX^i,t_jX^j)\big)}
where it is explicitly shown the realization through the CBH-formula.
Or in general from Eq. \siete\ the star-product is given by 

$$
{\rm exp}(s_iX^i) \star {\rm exp}(t_jX^j)
$$
\eqn\diezu{ 
= {\rm exp}(s_iX^i) \bigg ( 1 + \sum_{n=1}^{\infty}
{1 \over 2^n}  X^{k_1} f^{i_1j_1}_{k_1} \dots X^{k_n}f^{i_nj_n}_{k_n}
\buildrel{\leftarrow}\over{\partial}_{i_1} \dots \buildrel{\leftarrow}
\over{\partial}_{i_n}
\cdot 
 \buildrel{\rightarrow}\over{\partial}_{j_1} \dots 
 \buildrel{\rightarrow}\over{\partial}_{j_n} + \dots \bigg){\rm exp}(t_jX^j)}
where $\dots$ denotes the rest of the terms in the expansion of the exponential given in 
Eq. \cinco. For further details of the properties  of the CBH-quantization see \kathotia.

\vskip 1truecm

\subsec{Kontsevich's Quantization of ${\cal G}^*$}

Now we recall the basic formulas from the Kontsevich's construction of the
star-product on any Poisson manifold $M$ \k. 

The Kontsevich's formula is a sum over a suitable class of oriented labeled {\it
admissible} graphs ${\bf G}_n$ of order $n \geq 0$ (constructed from $n$ wedges graphs).
An admissible graph
$\Gamma$ is an element of ${\bf G}_n$ if it is constructed from $n+2$
{\it aerial} vertices $\{1,2,\dots , n \}$ and two {\it ground} fixed
vertices $\{L,R\}$ with $V_{\Gamma} = \{1,2, \dots , n\} \sqcup \{L,R\}$ and
$2n$ edges from the set of edges of $\Gamma$, $E_{\Gamma}$ such that for each vertex $v_k$ 
of $\{1,2, \dots,n \}$ there are a pair of edges emanating from $v_k$ to
any other of the aerial vertices except for $v_k$ itself {\it i.e.} there
are no loops. The pair of edges emanating from each $v_k$ of the aerial
vertices are labeled as $e^1_k$ and $e^2_k$. 

In \k\ Kontsevich found a correspondence between each graph $\Gamma$ 
of ${\bf G}_n$ and bi-differential operators in the form:

$$ \Gamma \in {\bf G}_n \Leftrightarrow B_{\Gamma, \alpha}:
C^{\infty}(M)  \otimes C^{\infty}(M) \to C^{\infty}(M)  $$ 
with $M$ the relevant`
Poisson manifold. Thus the Kontsevich star-product is 

\eqn\doce{ f \star_K g \equiv \sum_{n=0}^{\infty} \hbar^n \sum_{\Gamma \in {\bf G}_n}
W_{\Gamma} B_{\Gamma, \alpha}(f,g)}
with $W_{\Gamma}$ is given by

\eqn\trece{
W_{\Gamma} \equiv {1\over (2\pi)^n} \int_{{\cal H}_n} \bigwedge_{i=1}^n d\phi^h_{e^1_k}
\wedge d\phi^h_{e^2_k},}
where ${\cal H}_n$ is the space of configurations of $n$ numbered pairwise
distinct points on ${\cal H}$ given by the $n$-punctured upper half-plane 
${\cal H} = \{z\in \IC / Im(z) >0\}$ endowed with the Lobachevsky
metric. The integral determining the weights \trece\ it was proved  to
be absolutely convergent \k. Finally $\phi^h_{e^1_k}$ and $\phi^h_{e^2_k}$
are harmonic functions on ${\cal H}_n$ representing the angles between
$l(p,\infty)$ and the edges $e^1_k$ and $e^2_k$ respectively, measured 
counterclockwise from $l(p,\infty)$ and $B_{\Gamma, \alpha}(f,g)$ is given by

$$B_{\Gamma, \alpha}(f,g)= \sum_{I:E_{\Gamma} \to \{1,\dots , d \}}
\bigg[ \prod_{k=1}^n \bigg(\prod_{e \in E_{\Gamma}, e(*,k)} \partial_I (e) \bigg)
\alpha^{I(e^1_k)I(e^2_k)} \bigg]$$ 

\eqn\kont{ \times
\bigg( \prod_{e \in E_{\Gamma}, e(*,L)} \partial_I(e)\bigg)f \times
\bigg( \prod_{e \in E_{\Gamma}, e(*,R)} \partial_I(e)\bigg)g.}

In the particular case when the Poisson manifold $M$ is identified with ${\cal G}^*$, the
Kontsevich's formula has very interesting interconnections with the others deformation
quantizations of ${\cal G}^*$. 
It was shown in Refs. \refs{\kathotia,\dito,\sho} 
that taking the vector fields $X$ and $Y$ from Eq. \tres\ to be located  at the ground vertices
$L$ and $R$ of the Kontsevich's construction, the only possible graphs
associated  to the bi-differential operators $B_{\Gamma, \alpha}$, are of two types:
$(i)$ Loop graphs and $(ii)$ symmetric-admissible graphs. For the particular case of 
nilpotent Lie algebras, the loop graphs contribution is vanishing \kathotia\ and the 
Kontsevich's 
 deformation quantization star-product coincides exactly with the CBH star-product Eq. (2.11).
For more general Lie algebras (no necessarily nilpotent) loop graphs are non-zero but still
is possible to find a mapping providing the equivalence between the Kontsevich's and CBH
star-products \refs{\kathotia,\dito,\sho}.

\vskip 2truecm

\newsec{D-branes on Group Manifolds}

In this section we describe some elementary facts about D-branes in group manifolds which
we will need in the next section. Another description of D-branes in group manifolds in the
boundary state formalism would be relevant for the deformation quantization theory
\refs{\kato,\stanciu,\reck}. However
for the purposes of this paper we will follows the
lines of Refs. \refs{\severa,\gawedzki}.

It is well known that the propagation of 
closed strings in a group manifold ${\bf G}$ is described by  the WZW
Lagrangian for the field $g: \Sigma \to {\bf G}$ \refs{\gepner,\ed}
satisfy the closed string boundary conditions
$g(0,\tau) = g(2\pi,\tau)$

\eqn\funct{L(g)= -{k \over 8 \pi} \int_{\Sigma} d^2 \sigma  \sqrt{h}
h^{ij}
{\rm  Tr} \bigg( g^{-1} \partial_i g \cdot g^{-1}
\partial^j g \bigg) - ik \Gamma(g),}
where $h^{ij}$ is a metric on the closed Riemann surface $\Sigma$, Tr is an
invariant form on the Lie
algebra ${\cal G}$ of
${\bf G}$ and $\Gamma(g)$ is the Wess-Zumino term which is given by

\eqn\catorce{ \Gamma(g) = \int_B \tilde{g}^* \chi =
 {1 \over 12 \pi} \int_B  d^3 \sigma \varepsilon^{ijk}
{\rm Tr} \bigg(  g^{-1} \partial_i g  \cdot g^{-1} \partial_j g  \cdot g^{-1} 
\partial_k g \bigg).}
Here $B$ is a three-manifold which has as boundary $\Sigma$, {\it i.e.}
$\partial B = \Sigma$, and $\chi$ is the left and right invariant three form
on the group manifold ${\bf G}$ given by

\eqn\quince{\chi = {1 \over 12 \pi} {\rm Tr}\big( g^{-1} d g  \wedge g^{-1} dg  
\wedge g^{-1} d g\big).}
In the Eq. \catorce\ $\tilde{g}$ is an extension of $g$ to the volume $B$ {\it i.e.}
$\tilde{g}: B \to {\bf G}$. Such an extension exists if $\pi_2({\bf G})=0$. Finally the 
coupling constant $k$ is an integer-valued constant and it is an element of $H^3({\bf G},\IZ)$.
Actually the three-form $\chi$ on ${\bf G}$ is an integer-valued cohomology class, {\it
i.e.}
an element of $H^3({\bf G},\IZ)$. This implies that $\chi$ is a closed form on ${\bf G}$ but
$\chi$ never is the differential of a global two-form $\omega$ over ${\bf G}$. That means that
it is not possible to define a gobal two-form $\omega$ over ${\bf G}$.

For the case of open strings the Lagrangian \funct\ is not well defined. The reason of this is
as follows: the propagation of open strings in ${\bf G}$ generates an open
world-sheet $W$ on which it can be coupled to a global two-form $\omega$ on
${\bf G}$. For the reasons mentioned previously such a global form cannot be
defined on ${\bf G}$. Thus for open strings the first term of the rhs of the 
Lagrangian \funct\ is still well defined but the second one is not. 

Now in order to
consider a D-brane configuration in ${\bf G}$, that is, two parallel submanifolds
${\bf H}_1$ and ${\bf H}_2$ of ${\bf G}$ and open strings stretched between these
submanifolds \refs{\severa,\gawedzki}. One can associate locally the required two-form on each
submanifold ${\bf H}_i$ $i=1,2$ {\it i.e.} $C_i$ on ${\bf H}_i$ for $i=1,2$,
respectively. These forms satisfy

\eqn\dseis{ \chi|_{{\bf H}_i} = d C_i. }
That means that the $C_i$ forms are chosen such that their
exterior derivative  coincides with
the restriction of the three-form $\chi$ on each submanifold ${\bf H}_i$. 

Ambiguities of the WZ-term for this D-brane configuration is given by \severa\

\eqn\dsiete{\delta \Gamma(g)= \Gamma(g) - \int_{P_1} g^*(C_1) - \int_{P_2}
g^*(C_2).}
where $P_i$ is a two-dimensional submanifold of ${\bf H}_i$ for $i=1,2$ respectively
 and this is precisely the two-dimensional projection to ${\bf H}_i$ of the volume
generated
by the variation  of the word-sheet $W$ connecting ${\bf H}_1$ and ${\bf H}_2$
in the group manifold ${\bf G}$. 

Notice that this action is well defined only locally. Thus the fields $g$
are restricted to live in the corresponding submanifold ${\bf H}_i$ of ${\bf G}$. 
In general ${\bf H}_i$ are not necessary to be a subgroup of ${\bf G}$. However
the case in which it coincides with the conjugacy classes of ${\bf G}$
is quite interesting and it will be discussed in the next
section in the context of deformation quantization of adjoint and
coadjoint orbits of ${\bf G}$.

Thus WZW theory for open strings with their end points on the D-branes is well
defined in terms of the triple $(\chi, C_1, C_2)$. $m$ strings
propagating in ${\bf G}$ and satisfying the Dirichlet boundary conditions on the
group submanifolds ${\bf H}_i$ as above is given by

\eqn\docho{L_D(g)= -{k \over 4 \pi} \int_{W} d^2 \sigma 
{\rm  Tr} \bigg( g^{-1} \partial_z g \cdot g^{-1}
{\partial}_{\bar{z}} g \bigg) + {k\over 12 \pi i} \int_B \tilde{g}^* \chi -{k \over 4 \pi}
\sum_m \int_{W_m} \tilde{g}^*|_{W_m} C_m}
where $z$ and $\bar{z}$ are the complex coordinates on the disc $W$. The sum in Eq. \docho\
is over the number of components of the boundary on the disc $W$. Action \docho\ is well
defined and it is completely specified by the forms $(\chi, C_m)$. Such a pair can be interpreted 
as an element of the relative cohomology chain $H^*({\bf G}/{\bf H}_m, \IZ)$. The world-sheet
$W$ and in general higher-dimensional world-volumes $W_k$ of dimension $k$ can be interpreted as 
$k$-dimensional cycles  of the relative homology chain $H_*({\bf G}/{\bf H}_m, \IZ)$ {\it i.e.}
$k$-dimensional submanifolds of ${\bf G}$ whose boundary lies in the ${\bf H}_m$ D-branes. Thus
the relative integer homology classifies the different D-branes in the group manifold 
${\bf G}$. The $\IZ$-valued (co)homology is necessary in order that the Feynman  path integral
be well defined. 

Thus the WZW action for an open string in the D-brane configuration is given by

\eqn\dnueve{L_W(g)= -{k \over 4 \pi} \int_{W} d^2 \sigma 
{\rm  Tr} \bigg( g^{-1} \partial_z g \cdot g^{-1}
{\partial}_{\bar{z}} g \bigg) - {k \over 4 \pi} \int_W B,}
where $B= \tilde{g}(G)$ with $G$ is the local two-form defined on the image of $W$ under $g$, such
that $\chi = dG$.

If the boundary of $W$ is empty then the term $\int_W B$ is invariant under the
transformation $B \to B + d \Lambda$. But if the case is such that the boundary of
$W$ is no empty then the term $\int_W B$ will transforms as

\eqn\dnueveu{ \int_W B \to \int_W B + \int_{\partial W \cap {\bf H}} \Lambda.}
At this step one can imitate the standard case of D-branes in type II string
theories \branes\ by noting that $C-B$ on the D-brane is a closed form, then locally
we have  abelian degrees of freedom $A$ in the
boundaries of the open string attached to the D-brane and it is given by

\eqn\dnueved{ \int_{\partial W \cap {\bf H}} A.}
Thus the gauge invariance can be restored if the gauge transformation for the
$B$-field is accompanied by the transformation of $A$ in the form $A \to A -
\Lambda |_{\bf H}$. The gauge invariant field strength is not $F = dA$ but the
combination ${\cal B} = F -B$. Thus the effective action is

\eqn\dnuevet{L_W(g)= -{k \over 4 \pi} \int_{W} d^2 \sigma 
{\rm  Tr} \bigg( g^{-1} \partial_z g \cdot g^{-1}
{\partial}_{\bar{z}} g \bigg) - {k \over 4 \pi} \int_W {\cal B}.}
The gauge
field in two-dimensions has no-propagating degrees of freedom and it implies that
field strength ${\cal B}$ should be constant, ${\cal B} = {1 \over 2}
\varepsilon^{ij} {\cal B}_{ij}$.  Thus a ${\cal B}$-field term can be consistly
introduced into the WZW description of open strings in group manifolds when
D-brane configurations are included. In the next section we will show that
precisely this ${\cal B}$-field term will plays a crucial role in the derivation
of the Kontsevich's star-product for the dual Lie algebra ${\cal G}^*$.

In Ref. \yu\ it was shown that the explicit form of ${\cal B}$ can also be extracted from the
boundary condition in the closed string picture $J = - \bar{J}$ where $J$ and $\bar{J}$ are the
holomorphic and anti-holomorphic chiral currents of the WZW model
\refs{\kato,\stanciu,\reck,\yu} and it is given by

\eqn\bfield{ - {k \over 4 \pi} \int_W \varepsilon^{ij} {\rm Tr}\bigg( g^{-1} \partial_i g \cdot
{1 + Ad(g) \over 1- Ad(g)} \cdot g^{-1} \partial_j g \bigg)}
where $Ad(g)$ is the adjoint action of ${\bf G}$ on the Lie algebra ${\cal G}$. It is given by
$Ad(g)X := {d \over dt}|_{t=0} g(exp tX)g^{-1}$ for all $g$ in ${\bf G}$. In this picture, the
D-brane is identified with the (co)adjoint orbit (conjugacy class) of ${\bf G}$. This result
leads to a reconciliation between the approach \severa\ and the boundary state description
\refs{\kato,\stanciu}.

\vskip 2truecm


\newsec{D-branes in ${\bf G}$ and the Kontsevich Star Product on ${\cal G}^*$}

The aim of this section is the derivation of the Kontsevich's formula of
deformation quantization for the dual of a Lie algebra ${\cal G}^*$. We
will argue that this formula is encoded in the theory of D-branes on group
manifolds as given in Refs. \refs{\severa,\yu,\gawedzki} and reviewed in
the previous section.

First of all we shall recall that the effective WZW field theory on the D-brane 
in a group manifold ${\bf G}$ is given by the Eq. \dnuevet. For definiteness
we will take a compact and simple  
group manifold ${\bf G}$ whose dual Lie algebra 
${\cal G}^*$ be provided with the Kirillov-Poisson structure. 
It is an easy matter to show
that Eq. \dnuevet\ can be rewritten as

\eqn\vuno{L_W(g)= -{k \over 4 \pi} \int_{W} d^2 \sigma 
{\rm  Tr} \bigg( g^{-1} \partial_z g \cdot g^{-1}
{\partial}_{\bar{z}} g \bigg) - {1 \over \pi k} \int_{W} d^2 \sigma 
B_{ab} J^a(z) \bar{J}^b(\bar{z}),}
where $B_{ab}$ is an antisymmetric and constant two-form on ${\bf H}$ 
and  $J^a(z)$ and $\bar{J}^b(\bar{z})$ are the chiral currents of the WZW model
and they are vector fields in ${\cal G}$ satisfying 

\eqn\vtres{ J(z) = J^a T_a = - {1\over 2} k g^{-1} \partial_z g,}

\eqn\vcuatro{\bar{J}(\bar{z}) = \bar{J}^a T_a = - {1\over 2} k g^{-1} 
\partial_{\bar{z}} g,}
where $\{T_a\}$ with $a=1,\dots , dim({\bf G})$ is a basis of ${\cal G}$. It is well
known
from the theory of open strings that the disc $W$ can be transformed into the upper
half-plane ${\cal H}$ with the boundary of the disc mapped into the boundary of ${\cal H}$
located at $Im(z) =0$.

We consider the correlation functions on the disc $W$ of the operators
$g$ located in $0$ and $1$ in the boundary of $W$. These correlation functions will be defined taking only
the first term of the Lagrangian \vuno\ in the Feynman path integral

\eqn\vcinco{\langle g(1)\cdot g(0)\rangle = {1 \over Z}  \int {\cal D}g
\ g(1)\cdot g(0)  \ {\rm exp} 
\bigg( {k \over 4 \pi}
\int_W d^2 \sigma {\rm Tr} \big( g^{-1} \partial_z g \cdot g^{-1}
{\partial}_{\bar{z}} g \big) \bigg), }
where 

\eqn\vseis{Z = \int {\cal D} g  {\rm exp} \bigg( + {k \over 4 \pi}
\int_W d^2 \sigma {\rm Tr} \big( g^{-1} \partial_z g \cdot g^{-1}
{\partial}_{\bar{z}} g \big) \bigg).}

Since the fields $g$ are ${\bf G}$-valued mappings it is natural to assume that the insertion
 operators in $0$ and $1$ on the boundary of
the disc are given by

\eqn\vseist{ g(1) \equiv {\rm exp}\big(J \big), \ \ \ \ \ 
g(0) \equiv {\rm exp}\big(\bar{J} \big) }
with 

\eqn\vseisc{ J(z) \equiv J^a(z) T_a, \ \ \ \ \ \ \bar{J}(\bar{z}) =
\bar{J}^a(\bar{z})T_a, }
where $J^a(z)$ and $\bar{J}^a(\bar{z})$ are the holomorphic and anti-holomorphic
currents given in Eqs. \vtres\ and \vcuatro.

We propose  that the two-point correlation function for the complete action $L_W$ of
Eq. \vuno\ leads to the Kontsevich's formula of the deformation quantization of the dual
Lie algebra ${\cal G}^*$ of the group manifold ${\bf G}$. This implies that the mentioned Kontsevich's
formula is encoded in the theory of open strings 
with Dirichlet boundary conditions in group manifolds. Thus we should consider the 
complete action $L_W$ including the $B$-field term too. Coefficient of the
$B$-field term goes as $1/k$ and in the large-$k$ limit one can consider this
term  as a `perturbation' and use the usual $1/k$-perturbation theory. Before of proceeding
with the $1/k$-expansion we consider generically the effect of the $B$-field term in the 
two-point correlation functions.  
Thus one can define new (deformed) correlation functions $\langle g(1) \cdot g(0)\rangle^B$ for this
`perturbed' system in the form

\eqn\vsiete{\langle g(1)\cdot g(0)\rangle^B = {1 \over Z} \langle g(1)\cdot g(0)  
{\rm exp} \bigg( + {1 \over  \pi k} \int_{\cal H} d^2 \sigma 
B_{ab} J^a(z) \bar{J}^b(\bar{z}) \bigg) \rangle}
where now

\eqn\vocho{ Z = \int {\cal D} g  {\rm exp} \big( i L_W \big).}
In Eq. \vsiete\ we have replaced $W$ by ${\cal H}$. The point $(z,\bar{z})$ where
the holomorphic $J(z)$ and anti-holomorphic $\bar{J}(\bar{z})$ chiral currents are
evaluated, is a specific point of the interior of the ${\cal H}$-plane.

The product $g(1) \cdot g(0)$ inside the above correlation function is the usual group
product

\eqn\vochou{ g(1) \cdot g(0) = {\rm exp}\big(J \big) \cdot {\rm exp}\big( \bar{J}
\big).}
We can now apply the CBH-formula (2.1) to the above equation and it results

\eqn\vochod{{\rm exp}\big(J \big) \cdot {\rm exp}\big( \bar{J}\big) = {\rm exp}\big( H(J,\bar{J})\big),}
where

\eqn\vochot{H(J,\bar{J}) = J^a(z)T_a + \bar{J}^a(\bar{z})T_a + {1 \over 2}T_cf^c_{ab}
J^a \bar{J}^b + {1 \over 12}\big(T_ef^e_{ad} f^d_{bc}J^aJ^b \bar{J}^c +
T_e f^e_{dc} f^d_{ab} J^a\bar{J}^b \bar{J}^c \big) + \dots }
Here $\{ T_a \}$ $a=1,\dots , dim({\bf G})$ is a basis of ${\cal G}$ and $f^a_{bc}$ are the structure
constants of the Lie algebra of ${\bf G}$.
At this level one can consider the set of generators $T_a$ of the Lie algebra ${\cal
G}$ as local coordinate functions on ${\cal G}^*$ according to the correspondence (2.3). Thus
${\rm exp}(J^aT_a)$ and ${\rm exp}
(\bar{J}^a T_a)$ can be seen as functions on ${\cal G}^*$ {\it i.e.} as elements of
$C^{\infty}({\cal G}^*)$. It is very easy to remove the part $J + \bar{J}$ from the
CBH-formula \vochot\ and we get

\eqn\vochoc{ g(1) \cdot g(0) = D \hat{g}(1)\hat{g}(0),}
where $\hat{g}(1) = {\rm exp}(J^a(z)T_a)$ and $\hat{g}(0) = {\rm exp}(\bar{J}^a(\bar{z})T_a)$
are seen as functions on ${\cal G}^*$ and $D$ is the symbol
given by 

\eqn\vochos{ D= {\rm exp}\bigg({1 \over 2}T_cf^c_{ab}
J^a \bar{J}^b + {1 \over 12}\big(T_ef^e_{ad} f^d_{bc}J^aJ^b \bar{J}^c +
T_e f^e_{dc} f^d_{ab} J^a\bar{J}^b \bar{J}^c \big) + \dots \bigg).}
We can substitute this last expression into the deformed correlation function \vsiete\ and it yields

\eqn\vochoo{ \langle g(1)\cdot g(0)\rangle^B = {1 \over Z} \langle D \hat{g}(1) \hat{g}(0)
{\rm exp} \bigg( {1\over  \pi k} 
\int_{\cal H} d^2 \sigma 
B_{ab} J^a(z) \bar{J}^b(\bar{z}) \bigg) \rangle .}

In the large-$k$ limit one can perform the perturbative $1/k$-expansion 
of the $B$-field term where we shall recall that $B_{ab}$ must be constant by the
arguments outlined at the end of the Section 3. Thus we get

\eqn\vnueve{ \langle g(1)\cdot g(0)\rangle^B = {1 \over Z} \sum_{n=0}^{\infty} (+{1 \over
\pi k})^n
{1\over n!}
\int_{{\cal H}_n} dz_1 d\bar{z}_1 \dots dz_n d\bar{z}_n \langle D_n \hat{g}(1) \hat{g}(0)
\prod_{j=1}^n B_{a_j b_j} J^{a_j}(z) 
\bar{J}^{b_j}(\bar{z})\rangle , }
where ${\cal H}_n = \{(z_1, \dots , z_n) / z_i \in
{\cal H}, \ Im(z_j)> \epsilon, |z_j -z_i|>
\epsilon, i\not= j \}$ is a $2n$-dimensional real manifold,
$D_n$ is given by $D$ from Eq. \vochos\ evaluated in the $n$-th coordinate pair
$(z_n,\bar{z}_n)$.
The fact that $\hat{g}(0)$ and 
$\hat{g}(1)$ can be seen as smooth functions on  ${\cal G}^*$ it does not implies that
they can be seen as functions on the disc $W$. Actually they are operators and one has
to compute the 
remaining correlation functions on the disc. In order to do that one must
give the operator product expansion (OPE) of the operators $g$ with the 
currents $J$ and $\bar{J}$. They are given by \gepner\

\eqn\treinta{J^a(z) g(\xi) = -{1\over z} T^a g( \xi) + \dots }
and
\eqn\tuno{ \bar{J}^a(\bar{z}) g(\xi) = {1\over \bar{z}} g( \xi) T^a + \dots}
where $\dots$ denotes the non-singular terms of the OPE. 
Applying these equations one can shown that the Eq. \vnueve\ turns out to be

$$\langle g(1) \cdot g(0)\rangle^B = {1 \over Z} \sum_{n=0}^{\infty}(-{1 \over  \pi k})^n 
{1 \over n!}
\int_{{\cal H}_n} dz_1 d\bar{z}_1 \dots dz_n d\bar{z}_n \ 
\prod_{k=1}^n  {1 \over z_k \bar{z}_k} $$

\eqn\tcuatro{\times \  \langle B_{a_1b_1} \dots B_{a_nb_n}
T^{a_1} T^{b_1}  \dots T^{a_n} T^{b_n} D_n \hat{g}(1)\hat{g}(0) \rangle .}
In terms of the bi-differential operator $\hat{\cal D}$ from Eq. (2.6) we get 

\eqn\tcuatrou{\langle g(1) \cdot g(0)\rangle^B = \sum_{n=0}^{\infty}(-{1 \over  \pi k})^n 
{1 \over n!}
\int_{{\cal H}_n} dz_1 d\bar{z}_1 \dots dz_n d\bar{z}_n \
\prod_{k=1}^n  {1 \over z_k \bar{z}_k} 
\langle\hat{\cal D}_n\big(\hat{g}(1), \hat{g}(0)\big)\rangle ,}
where we have absorbed the terms with $B$'s and $T$'s in the normalization factor. Thus
the above formula can be written as

\eqn\tcinco{\langle g(1) \cdot g(0)\rangle^B = \sum_{n=0}^{\infty} (-{1  \over \pi k})^n 
\sum_{\Gamma}
 W_{\Gamma} \langle\hat{\cal D}_{\Gamma}\big(\hat{g}(1),\hat{g}(0)\big)\rangle,}
where 
\eqn\tsiete{ W_{\Gamma} = {1 \over n!} \int_{{\cal H}_n} dz_1 d\bar{z}_1 \dots 
dz_n d\bar{z}_n \
\prod_{k=1}^n  {1 \over z_k \bar{z}_k}}
with $\Gamma$ representing an admissible graph constructed from vertices and edges following
the recipe of subsection 2.3.

Thus one can see that the correlation function $\langle g(1)\cdot g(0)\rangle^B$ looks like the 
Kontsevich's product for the Lie algebra ${\cal G}^*$ which is given by

\eqn\tcincou{\hat{g}(1) \star_K \hat{g}(0) \equiv \langle g(1) \cdot g(0)\rangle^B = 
\sum_{n=0}^{\infty} (-{ 1 \over \pi k})^n
\sum_{\Gamma \in {\bf G}_n} W_{\Gamma} B_{\Gamma}(\hat{g}(1),\hat{g}(0))}
where we have identified $\hat{\cal D}_{\Gamma}(\hat{g}(1),\hat{g}(0))$ to  $B_{\Gamma}
(\hat{g}(1),\hat{g}(0))$ just as one would expect. 

Formula \tcincou\ reproduces exactly the Konsevich's star-product Eq. (2.12) defined on the
space of smooth functions on the dual Lie algebra ${\cal G}^*$ of the group manifold ${\bf G}$. 
It is remarkable that the Kontsevich's formula was obtained explicitly as a $1/k$-expansion
for the large-$k$ limit. This is a typical characteristic of a string perturbative expansion in its
string coupling constant in the correspondence between the topological sigma model for open
strings and the Chern-Simons effective field theory \cs. 
In fact, the close relation between Chern-Simons and WZW model leads us to think that a
description of the Kontsevich's formula in terms of the Feynman diagrams of the Chern-Simons
theory is possible. Actually it is no surprising form the results of
\refs{\cattaneo,\schwarz,\schomerus}.

Several comments are in order. The first one concerns with the relation with the
boundary states description of D-branes in group manifolds \refs{\kato,\stanciu,\reck}.
From there it is known that Dirichlet boundary conditions for open strings in groups
can be encoded in the algebraic structure of the corresponding superconformal algebra.
Within this prescription it was shown in \yu\ that for D-branes ${\bf H}$ with group
structure, {\it i.e.} subgroups of ${\bf G}$, for instance adjoint or coadjoint orbits
${\bf C}$ of ${\bf G}$. Boundary state prescription implies that open strings can end
on this orbits ${\bf C}$ defining D-particles.  From \stanciu\ if ${\bf C}$ constitutes
a Lagrangian submanifold of ${\bf G}$ then $T{\bf G}|_{\bf C} = T {\bf C} \oplus N {\bf
C}$, where $T {\bf C}$ and $ N {\bf C}$ are the respective tangent and normal bundle to
${\bf C}$. If we have a distinguished point (the identity) it turns out to be ${\cal G}
= {\cal C} \oplus {\cal C}^{\bot}$ at the level of Lie algebras. Now if we assume that
the boundary components of the disc are mapped into the real part of ${\cal G}$ {\it
i.e.} are mapped to ${\cal C},$ then the fields of the Lagrangian \vuno\ will be ${\cal
C}$-valued fields and therefore the deformation quantization formula Eq. \tcincou\ will
describe the deformation quantization of the coadjoint orbits ${\bf C}$. In
particular, for semi-simple Lie groups ${\bf G}$, the corresponding star-product fom
the CBH-quantization should gives the quantization described in \fioresi.

Following Ref. \vafa\ it is well known that a topological field theory should lives 
in the corresponding D-brane wrapped in the Lagrangian submanifolds of ${\bf G}$. 
This topological field theory is
provided by the effective field theory on the D-brane $W$ given by the WZW Lagrangian for open
strings or its extensions to gauged WZW-theory \refs{\stanciu,\kz}. It is expected that a
deeper treatment of the BRST and BV approach of WZW theory of D-branes would be a better
framework to
understand the Kontsevich's formula for ${\cal G}^*$ and its description in terms of the
formality conjecture in a
similar way than Cattaneo and Felder
\cattaneo\ found the Kontsevich's formula from a topological non-linear sigma model for open
strings by using the BV quantization method. We leave this subject for a future investigation.

\vskip 2truecm

\newsec{Concluding Remarks}

The close relation between deformation quantization theory (and in general 
noncommutative geometry) and the theory of D-branes has appeared in diverse 
ways in the literature 
\refs{\cattaneo,\schwarz,\maxim,\schomerus,\douglas}. In the present 
paper we found new evidence which favored this relation. Here we have explored 
some interplay between the D-brane theory on the group
manifold ${\bf G}$ and the Kontsevich's star-product on the dual Lie algebra
${\bf G}^*$ of the group ${\bf G}$. Our derivation 
has some similarity to the derivation of the Kontsevich's product on $\IR^n$ by
using the two-point correlation functions from the perturbative theory of open 
strings in $\IR^n$ in a background $B$-field \schomerus. However various
different things are in order. In our case the relevant theory is the effective 
WZW-field theory describing D-branes living in the group manifold ${\bf G}$ \severa. 
This effective 
field theory is given by the pairing in relative (co)homology chain $H({\bf G}/{\bf H},
\IZ)$. This implies the consistent existence of a background $B$-field term which is
coupled to the world-sheet $W$. This $B$-field term, as we have 
showed, it generates a $1/k$-expansion which can be (under some specifications)
identified with the Kontsevich's formula for ${\cal G}^*$. 

It is interesting  to conclude from our results that the Kontsevich's formula
on ${\cal G}^*$ given Eq. \tcincou\ can be seen as a $1/k$-perturbative expansion of the 
WZW effective field theory on a D-brane in ${\bf G}$.
Moreover Witten has showed in
\cs\ that Chern-Simons perturbation theory can been seen as a perturbative open
string theory with the $1/k$-expansion of the Chern-Simons theory coinciding with
the perturbative expansion of the string theory. Chern-Simons field theory represents the
low energy effective field theory of the string theory with the string coupling constant given by
$1/k$. Thus our result of getting the Kontsevich's formula as a perturbative expansion in the
string coupling constant confirms the current statement that the deformation quantization of any
Poisson manifold needs from stringly concepts \refs{\cattaneo,\schomerus,\maxim}.

It is very well known that 
$1/k$-perturbative expansions in WZW theory are related to $1/k$-perturbative expansions in the 
Chern-Simons theory. Thus one can, in principle, to translate our results in terms of a 
topological perturbative invariants of Chern-Simons theory. It would be interesting to 
study the Kontsevich's formula as given in \cattaneo\ and to look for its relation to the
Rozansky-Witten invariants \roza. 
It would be interesting to explore also further interplay between the Poisson-Lie
T-duality of the WZW theory for D-branes \severa\ and the relevant algebraic 
relations involved in the quantization of Lie bi-algebras as worked out in \kaz. Also, following
\schomerus\ one would trying to generalize the CBH-quantization and its relation to Kontsevich's
formula  
to curved manifolds (Riemannian phase spaces) through the application of the boundary deformation
theory \recknagel. It would be interesting to compare these results to those obtained by 
Bordemann et al \bordemann\ in the context of Fedosov's formalism.

Finally, the search for a deeper interplay between the string theory tools as: conformal
field, theory, Maldacena's conjecture,  D-branes and D-branes in group manifolds, moduli
spaces etc. and
mathematics tools as: formality conjecture, operads,
motives etc. pointed out in \maxim\ deserves intensive further investigation.


\vskip 2truecm

\centerline{\bf Acknowledgements}
One of us (H. G.-C.) would like to thank Prof. E. Witten for
his hospitality at
the Institute for Advanced Study. It is a
pleasure to thank I. Carrillo for useful discussions.



\listrefs

\end